\documentclass{desyproc}

\usepackage[T1]{fontenc}
\usepackage{enumitem}
\setenumerate[1]{label=[\arabic*]}

\usepackage{natbib}

\begin{document}
\title{Rare Low-Energy Event Searches with the \textsc{Majorana Demonstrator}}

\author{{\slshape Gulden Othman$^{1,2}$, \textit{on behalf of the \textsc{Majorana} collaboration}}\\[1ex]
$^1$University of North Carolina at Chapel Hill, NC, USA; $^2$Triangle Universities Nuclear Laboratory, Durham, NC, USA}

\contribID{Othman\_Gulden}

\confID{20012}  
\desyproc{DESY-PROC-2018-03}
\acronym{Patras 2018} 
\doi  

\maketitle

\begin{abstract}
The \textsc{Majorana Demonstrator} is currently searching for neutrinoless double-beta decay in $^{76}$Ge and will demonstrate the feasibility to deploy a tonne-scale experiment in a phased and modular fashion. It consists of two modular arrays of natural and $^{76}$Ge-enriched germanium detectors totaling 44.1 kg, of which 29.7 kg is enriched, located at the 4850' level of the Sanford Underground Research Facility in Lead, South Dakota, USA. The low-backgrounds and low trigger thresholds ($<$1 keV) achieved by the \textsc{Demonstrator} allow for additional rare-event searches at low-energies, e.g. searches for WIMPs, bosonic dark matter, and solar axions. In this work, we will present results and ongoing efforts related to these rare-event searches and discuss the future reach of \textsc{Majorana}. 
\end{abstract}

\section{The \textsc{Majorana Demonstrator}}
The \textsc{Majorana Demonstrator} [1]
is a search for neutrinoless double-beta decay ($0\nu\beta\beta$) in $^{76}$Ge currently operating at the 4850' level of the Sanford Underground Research Facility (SURF) in Lead, SD, USA. The goals of the \textsc{Majorana Demonstrator} are to demonstrate backgrounds low enough to justify building a tonne-scale experiment, to demonstrate the feasibility to construct modular arrays of high-purity Ge (HPGe) detectors, and for additional searches for physics beyond the Standard Model (SM). The \textsc{Majorana Demonstrator} employs 44.1 kg P-type point contact (PPC) HPGe detectors, 29.7 kg of which is enriched to 88\% $^{76}$Ge. The detectors are arranged into two separate cryostats constructed of ultra pure underground electroformed copper (UGEFCu). Starting from the innermost region, two cryostat modules are surrounded by an inner layer of electroformed copper, an outer layer of commercial copper, lead, an active muon veto, borated polyethylene, and polyethylene.

The \textsc{Majorana Demonstator} is currently operating with the best energy resolution of any $0\nu\beta\beta$ experiment with an energy resolution of 2.5 keV FWHM at 2039 keV, the Q-value for $0\nu\beta\beta$ in $^{76}$Ge. Recent results released at the Neutrino 2018 conference include a blind analysis of data with 26 kg-yr enriched exposure (11.85 kg-yr blind data). An observed background rate of 15.4 $\pm$ 2.0 cts/FWHM t yr was found near the $^{76}$Ge Q-value for a full exposure limit on the half life of $T^{0\nu}_{1/2} >2.7\times10^{25}$ [2, 3]. 

\section{The \textsc{Majorana} Low-Energy Program}

The low backgrounds and detector thresholds achieved by the \textsc{Majorana Demonstrator} allow for additional searches for physics beyond the SM at low-energies ($<$100 keV). These include searches for bosonic dark matter, Pauli exclusion principle violation, electron decay, solar axions [4], and light ($\sim$10 GeV/c$^2$) WIMPs. Careful material selection and handling, as well as extreme care taken to reduce the amount of cosmogenic activation in the enriched detectors results in low-backgrounds down to low-energies. Additionally, the low-capacitance achieved by PPC detector technology coupled with the development of a low-noise, low-mass front-end [5] enable low trigger thresholds ($<$1 keV) and excellent energy resolution (0.4 keV FWHM at 10.4 keV). Excellent pulse-shape analysis abilities enable further background discrimination. 

\label{lowEspectrum}
\begin{figure}[hb]
\centerline{\includegraphics[width=0.55\textwidth]{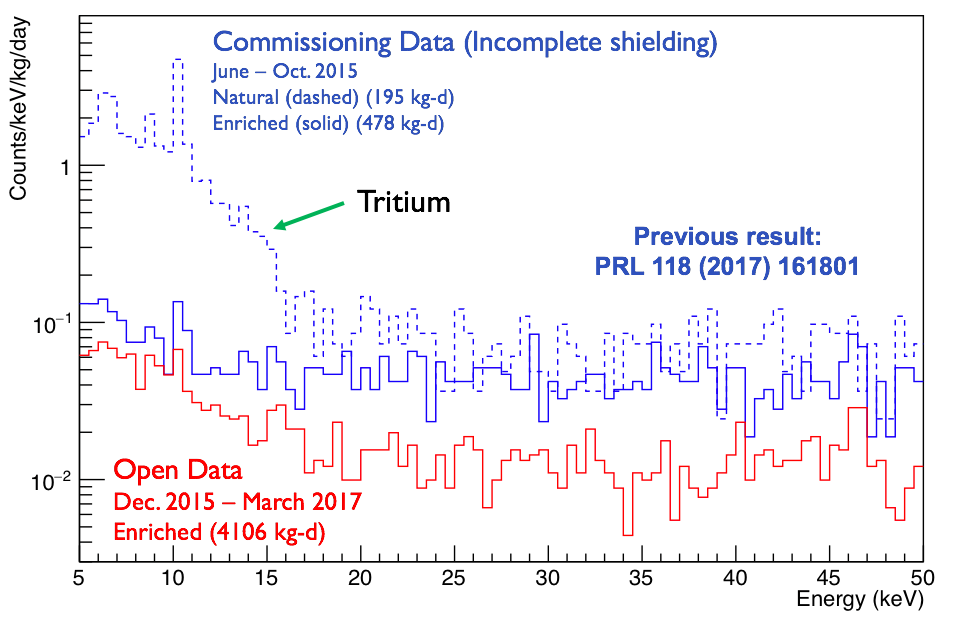}}
\caption{Spectrum from 5-50 keV of data from the \textsc{Majorana Demonstrator}. Data shown solid (dashed) blue is from commissioning data before shielding was completed from enriched (natural) detectors. Data in red is after the addition of the inner electroformed copper shield.}\label{fig:lowEspec}
\end{figure}

\section{Previous Results}

The \textsc{Majorana} background spectrum from 5-50 keV is shown in Fig. 1. There is a clear difference visible in the enriched and natural detectors due to the fact that, unlike the enriched detectors, the natural detectors had a much longer surface exposure, resulting in increased cosmogenic activation.

\begin{figure}
\begin{center}
\includegraphics[width=0.55\textwidth]{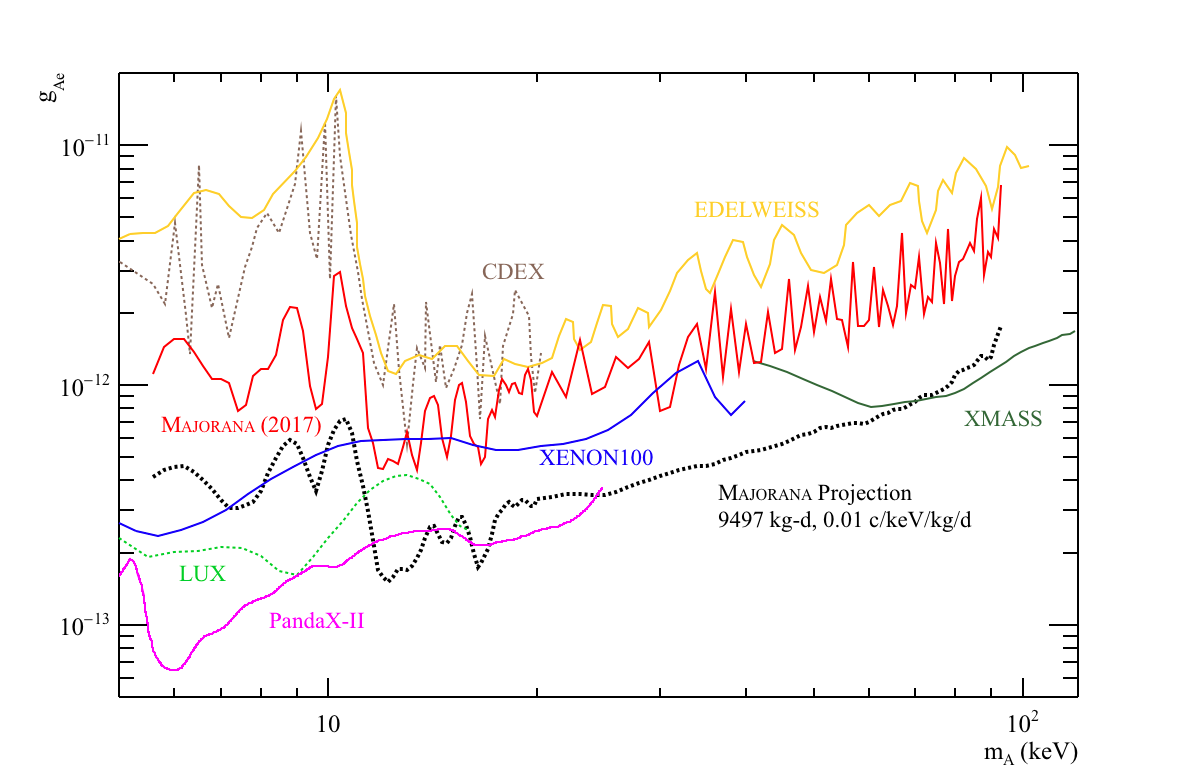}\caption{Current \textsc{Majorana} limits (red) and sensitivity projections based on exposure (dotted black) for pseudoscalar bosonic dark matter.}\label{fig:alps}
\end{center}
\end{figure}

The \textsc{Majorana} collaboration has set limits on bosonic dark matter [4]. This is accomplished by searching for peaks at the mass of the dark matter particle in the spectrum, which is assisted by the excellent energy resolution of the \textsc{Majorana} detectors. The spectrum in solid blue in Fig. \ref{fig:lowEspec} was used to set the previous limits on pseudoscalar bosonic dark matter shown as the solid red line in Fig. \ref{fig:alps}. The dotted black line shows the projected sensitivity using new data with 9497 kg-d of exposure including blind data, assuming a background rate of 0.01 cts/keV/kg/d. 


\begin{figure}
\begin{center}
\includegraphics[width=0.55\textwidth]{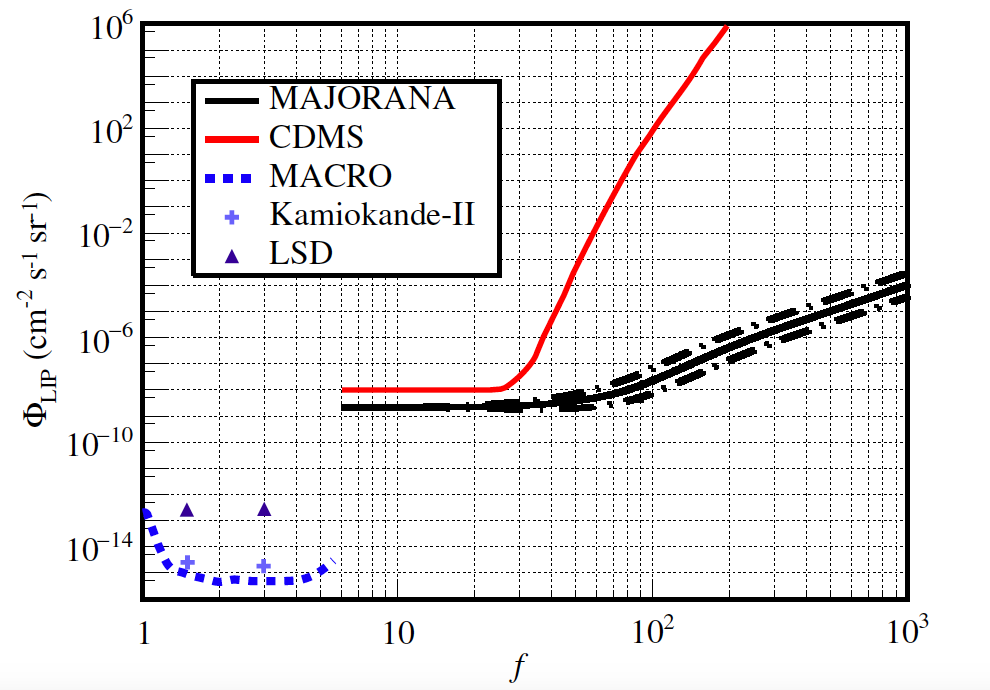}\caption{The \textsc{Majorana} 90\% confidence interval (solid black) and 1$\sigma$ uncertainty bands (dashed black) for Lightly Ionizing Particles [6].}\label{fig:lips}
\end{center}
\end{figure}

In addition to dark matter, the \textsc{Majorana} collaboration has set limits on lightly-ionizing particles (LIPs). LIPs are theoretical particles whose electromagnetic interactions are suppressed compared to the normal interaction strength. This would manifest experimentally as particles which have a fractional charge compared to the fundamental charge. Limits are typically set based on a reduction factor \textit{f}. The charge $q$ detected is related to the reduction factor \textit{f} by $q = e/f$, where $e$ is the fundamental charge. The \textsc{Majorana Demonstrator} completed a background free search for LIPs resulting in the limit shown in Fig. \ref{fig:lips} [6].

\section{Ongoing Analyses} 

 In addition to the results shown in the previous section, the \textsc{Majorana} collaboration is working on pulse-shape analysis efforts to further reduce backgrounds at low-energies in order to improve the analysis thresholds. Lower analysis thresholds would enable a host of additional searches at low energies, including searches for solar axions and low-mass WIMPs. 

 The \textsc{Majorana} collaboration is able to search for solar axions through two different coupling mechanisms: the Primakov effect as well as axio-electric effect. Solar axions can be detected via the Coherent Primakoff effect, even without exact knowledge of the directions of the crystal axes, if there are more than about 15 detectors [7]. In this case, the true total axion rate can be approximated by averaging angle-specific rates over all possible angles. This would manifest experimentally as a time-dependent spectrum with signals expected between 3-10 keV. Additionally, solar axions can be detected via the axio-electric effect [8]. This is analogous to the photoelectric effect, in that an axion is absorbed and the energy is released as electrons. This would manifest as characteristic peaks in a spectrum below 15 keV. 

 The \textsc{Majorana Demonstrator} is also sensitive to low-mass WIMPs ($\sim$10 GeV/c$^2$). Though the \textsc{Demonstrator} is unable to distinguish between events originating from nuclear recoils and those due to electron recoils, the low intrinsic backgrounds will enable a search which can provide an important check for experiments searching for WIMPs using Ge. 
\section*{Acknowledgments}
 This material is based upon work supported by the U.S. Department of Energy, Office of Science, Office of Nuclear Physics, the Particle Astrophysics and Nuclear Physics Programs of the National Science Foundation, and the Sanford Underground Research Facility. This work is also funded by the National Science Foundation Graduate Research Fellowship Program under grant number DGE-1144081. 

 \section*{References}

\noindent \begin{enumerate}[itemsep=0.07mm, align=left] 
 \item {N. Abgrall et al. The \textsc{Majorana Demonstrator} Neutrinoless Double-Beta Decay Experiment. \textit{Adv. High Energy Phys.}, 2014:365432, 2014.}

 \item {C. E. Aalseth et al. Search for Neutrinoless Double-$\beta$ Decay in $^{76}$Ge with the \textsc{Majorana
Demonstrator}. \textit{Phys. Rev. Lett.}, 120(13):132502, 2018.} 

\item {Vincente Guiseppe. New Results from the \textsc{Majorana Demonstrator} Experiment.
\textit{Zenodo}, http://doi.org/10.5281/zenodo.1286900, June 2018.}

\item {N. Abgrall et al. New limits on bosonic dark matter, solar axions, pauli exclusion principle
violation, and electron decay from the \textsc{Majorana Demonstrator}. \textit{Phys. Rev. Lett.}, 118:161801,
Apr 2017.} 

\item {P. Barton et al. Low-noise low-mass front end electronics for low-background physics experiments
using germanium detectors. In \textit{2011 IEEE Nuclear Science Symposium Conference
Record}, pages 1976-1979, Oct 2011.} 

\item {S. I. Alvis et al. First limit on the direct detection of lightly ionizing particles for electric
charge as low as $e/1000$ with the \textsc{Majorana Demonstrator}. \textit{Phys. Rev. Lett.}, 120:211804,
May 2018.}

\item {Wenqin Xu and Steven R. Elliott. Solar axion search technique with correlated signals from
multiple detectors. \textit{Astroparticle Physics}, 89:39 - 50, 2017.} 

\item {Javier Redondo. Solar axion flux from the axion-electron coupling. \textit{JCAP}, 1312:008, 2013.} 

\end{enumerate}


\end{document}